\documentclass{webofc}
\usepackage[varg]{txfonts}   
\def\bfg #1{{\mbox{\boldmath $#1$}}}
\begin{document}
\title{Search for time-reversal-invariance violation in double polarized 
antiproton-deuteron scattering}
%
%

\author{\firstname{Yuriy} \lastname{Uzikov}\inst{1,2,3}\fnsep\thanks{\email{uzikov@jinr.ru
    }}
   \and \firstname{Johann} \lastname{Haidenbauer}
   \inst{4}\fnsep\thanks{\email{j.haidenbauer@fz-juelich.de
    }}
}
\institute{ {Joint Institute for Nuclear Research, Dubna 141980, Russia }
\and
          { Dubna State University, Dubna 141980, Russia}
\and
   { Department of Physics, M.V. Lomonosov Moscow State University, Moscow 119991, Russia}         
     \and     
{Institute for Advanced Simulation and Institut f\"ur Kernphysik, Forschungszentrum J\"ulich
GmbH, D-52428 J\"ulich, Germany} 
          }
 
\abstract{

  Apart from the $pd$ reaction also the scattering of antiprotons with transversal 
  polarization $p_y^p$ on deuterons with tensor polarization $P_{xz}$ provides 
  a null-test signal for time-reversal-invariance violating but parity 
  conserving effects.
  Assuming that the time-reversal-invariance violating 
  $\bar NN$ interaction contains the same operator structure as the $NN$ interaction,
  we discuss the energy dependence of the null-test signal in $\bar pd$ scattering
  on the basis of a calculation within the spin-dependent Glauber theory at 
  beam energies of 50-300 MeV.
}
\maketitle
\section{Introduction}
\label{intro}

 Under CPT symmetry time-reversal-invariance violating but parity conserving 
 (TVPC) forces are considered as a possible source of CP-invariance violation, which is 
 required to account for the matter-antimatter asymmetry in the universe \cite{sakharov}. 
 In contrast to effects from time-reversal-invariance 
 violation together with parity violation such as a permanent electric dipole moment (EDM) of 
 elementary particles, so far much less attention was paid to TVPC effects. 
 The reason why TVPC effects are interesting  
 is that experimental limits on them are still rather weak, in particular, considerably
 weaker than those for the EDM.
      
      Since the intensity of TVPC interactions within the standard model is extremely small 
      \cite{conti-khripl},
      an observation of any effects at the present accuracy level of experiments 
      would be a direct indication of physics beyond the standard model.
      Indeed a pertinent measurement is planned at the COSY accelerator in the Research Center
in J\"ulich \cite{TRIC}. The observable in question is the integrated cross section for scattering of 
      protons with transversal polarization $p_y^p$ on deuterons with tensor polarization $P_{xz}$.
      It provides a null-test signal for TVPC
      effects \cite{conzett} and it will be measured in $pd$ scattering at 135 MeV \cite{TRIC}.
      Theoretical studies of the energy dependence of the expected signal were performed at energies
      of the planned experiment \cite{Beyer,Lazauskas,UZTemPRC,Uzdspin15,UZEPJweb,UzTemIJMF2016,
      UZJHPRC16,UZEPJweb17} 
 on the basis of the spin-dependent Glauber theory and demonstrate several unexpected effects. 
 Among them are (i) the absense of the contribution from 
 the lowest-mass meson-exchange ($\rho$ meson) in the TVPC $NN$ interaction,
 caused by its specific isospin, spin and momentum dependence; 
 (ii) a strong impact of the deuteron $D$-wave on the null-test signal
 due to a destructive interference between the $S$- and $D$-wave contributions, 
 even for zero transferred 3-momentum; (iii) oscillating behaviour 
 of the null-test signal as a function of the beam energy, i.e. the vanishing of the TVPC signal 
 at some specific energies is possible even when the TVPC interaction itself is nonzero; 
 (iv) a very small influence of the Coulomb interaction on the TVPC term of the $pd$ forward 
scattering amplitude $\widetilde {g}$. 
 Furthermore, certain relations between differential observables of elastic $pd$ scattering
 caused by time-reversal-invariance requirements were obtained
 and the degree of their violation by TVPC $NN$ forces was studied~\cite{TUZizv,TUZizv16}. 
   
   Since the spin structure of the amplitude for $pd$- and $\bar p d$ elastic scattering 
   is the same, it is obvious that the integrated cross section for scattering of a
   polarized ($p_y^{\bar p}$) antiproton on tensor polarized ($P_{xz}$) deuterons also 
   provides a null-test signal for TVPC effects. Furthermore, the TVPC $\bar NN$ amplitude for 
 elastic scattering contains the same operator structures as 
   the one for TVPC $NN$ elastic scattering, except for the charge-exchange terms.
    Therefore, the formalism developed in Refs.~\cite{UZTemPRC,Uzdspin15,UZJHPRC16} within
    the Glauber theory for the calculation of the null-test
    signal in $pd$ scattering can be straightforwardly applied to $\bar p d$ scattering too. 
    However, due to differences in the hadronic part of the $pN$ and $\bar p N$ scattering 
    amplitudes and also in the electromagnetic interactions, 
    the energy dependence of the null test signal in $pd$ and
    ${\bar p} d$ interaction has to be different. 
    In the present work the energy dependence of the null-test signal in ${\bar p} d$ scattering
    is  studied on the basis of calculations within the spin-dependent Glauber theory
    using the spin-dependent $\bar p N $ amplitudes from a recent partial wave
analysis of $\bar pp$ scattering \cite{Nijmegen}.

\section {Null-test signal for time-reversal-invariance violation}
\label{sec-1}

 The total cross section for $\bar pd$ scattering with TVPC forces included 
 can be written in the same form as for $pd$ scattering \cite{UZTemPRC}
 \begin{equation}
\label{totalspin}
{
\sigma_{tot}= {\sigma_0^t+\sigma_1^t{{\bf p}^{\bar p}\cdot {\bf p}^d}+
 \sigma_2^t {({\bf p}^{\bar p}\cdot {{\bf m}}) ({\bf p}^d\cdot { {\bf m}})}+
\sigma_3^t { P_{zz}}} +{\widetilde \sigma} {p_y^{\bar p} P_{xz}^d} \, .
}
\end{equation}
Here ${\bf p}^{\bar p}$  (${\bf p}^d$) is the
vector polarization of the  initial antiproton (deuteron),
$P_{zz}$ and $P_{xz}$ are the tensor polarizations of the deuteron, and 
$p_y^{\bar p}$ is the transversal component of the antiproton vector polarization. 
The OZ axis is directed along the beam direction
 ${{\bf m}}$, the OY axis is directed along the vector polarization of
 the  antiproton beam ${\bf p}^{\bar p}$
 and the OX axis is chosen to form a right-handed reference frame. The integrated cross 
 sections $\sigma_i^t$ ($i=0,1,2,3$) are those which arise from a standard 
time-reversal invariant and parity conserving interaction, while the 
 last term ${\widetilde \sigma}$  appears only in the presence of the TVPC interactions
 and constitutes the TVPC null-test signal. The result (\ref{totalspin}) can be derived using
 phenomenological $\bar p d$ forward scattering amplitudes and the generalized optical theorem. 

The evaluation of the integrated cross sections $\sigma_i^t$  and 
$\widetilde {\sigma}$ at beam energies $>100$ MeV
can be done on the basis of the spin-dependent Glauber theory of ${\bar p}d$ scattering 
which is formulated similarly to the theory of $pd$ scattering given in Ref.~\cite{PK}.
Indeed, as  shown in Ref.~\cite{TUZyaf}, this theory allows one to describe rather well 
available data on differential spin observables of $pd$ scattering in the forward 
hemisphere at beam energies of $135-200$ MeV. 
For the antiproton-deuteron scattering this theory can be applied at even lower 
energies due to the presence of strong annihilation effects.
In the  Glauber theory one uses the elastic (on-shell) ${\bar N}N$ scattering amplitudes 
as input. Hadronic amplitudes of the ${\bar p}N$ scattering are taken here in the same 
form as for $pN$ scattering
 \cite{PK}
\begin{eqnarray}
\label{pnamp}
M_N({\bf p}, {\bf q};\bfg \sigma, {\bfg \sigma}_N)=
 A_N+C_N\bfg \sigma \hat{\bf  n} +C_N^\prime\bfg \sigma_N \hat{\bf  n }+
B_N(\bfg \sigma \hat {\bf k}) (\bfg \sigma_N \hat {\bf k})+\\ \nonumber
+ (G_N+H_N)(\bfg \sigma \hat {\bf q}) (\bfg \sigma_N \hat {\bf q})
+(G_N-H_N)(\bfg \sigma \hat {\bf n}) (\bfg \sigma_N \hat {\bf n}) \, ,
\end{eqnarray}
where  ${\hat {\bf q}}$, ${\hat {\bf k}}$ and ${\hat {\bf n}}$
 are defined as unit vectors along the vectors  ${ {\bf q}}=({\bf p}-{\bf p}')$,
${ {\bf k}}=({\bf p}+{\bf p}')$
and ${ {\bf n}}=[ {\bf k}\times {\bf q}]$,
respectively; ${\bf p}$ (${\bf p}'$) is the initial (final) antiproton momentum.

In general, the TVPC $NN$ interaction contains 18 different terms~\cite{herczeg}. 
In the case of the on-shell $NN$ scattering amplitude there are only 
three terms with different (independent) spin-momentum structures.
In the present study we consider the
following two terms for the TVPC (on-shell) $t$-matrix of elastic ${\bar p}N$ scattering
which have the same structure as those in TVPC $pN$ scattering
\begin{eqnarray}
\label{TVbNN}
t_{{\bar p}N}=
{h_N[({\bfg \sigma} \cdot {\bf k})({\bfg \sigma}_N \cdot {\bf q})+
({\bfg \sigma}_N \cdot {\bf k})({\bfg \sigma} \cdot {\bf q})-
\frac{2}{3}({\bfg \sigma}_N \cdot{\bfg \sigma})
({\bf k}\cdot {\bf q}) ]}/m_p^2
+ \\ \nonumber
+g_N [{\bfg \sigma} \times {\bfg \sigma}_N]\cdot [{\bf q }\times{\bf k}]
[{\bfg \tau} -{\bfg \tau}_N]_z/m_p^2.
\end{eqnarray}
Here ${\bfg \sigma}$ (${\bfg \sigma}_N$) is the Pauli matrix acting on the spin 
state of the antiproton (nucleon $N=p,n$) and ${\bfg \tau}$ (${\bfg \tau}_N$)
 is the isospin matrix acting on the isospin state of the antiproton (nucleon). 
The momenta $\bf q$ and $\bf k$ were already defined above in the context of 
Eq.~(\ref{pnamp}).
Both terms in Eq.~(\ref{TVbNN}), $h_N$ and $g_N$, occur in the TVPC $pn$ interaction. 
The TVPC $pN$ scattering amplitude contains also the charge-exchange term
 \begin{eqnarray}
  \label{chargex}
t^{ch}=  {g_N^\prime ({\bfg \sigma} - {\bfg \sigma}_N)\cdot i\,[{\bf q}\times {\bf k}]
[{\bfg \tau} \times{\bfg \tau}_N]_z}/m_p^2,
 \end{eqnarray}
which describes the elastic transitions $pn\to np$ and $np\to pn$.  
Within a picture of one-meson-exchange interaction this $g^\prime$-term
corresponds to the charged $\rho$-meson exchange \cite{simonius75}.  
 The same term (\ref{chargex}) corresponds to the charge-exchange processes $\bar p p\to \bar n n$
 or $ \bar n n \to \bar p p$. However, in contrast to $pn$ scattering
 these processes  are inelastic and therefore the operation of
 time-reversal invariance transforms, for example, the $\bar p p\to \bar n n$ amplitude to
 the $\bar n n \to \bar p p$ amplitude and does not impose any restrictions on
 these amplitudes. 
 
 The $h_N$-term in Eq.~(\ref{TVbNN}) can be associated with the axial $h_1$-meson exchange.
 As shown in Ref. \cite{simonius75}, contributions of the $\pi$- and $\sigma$-meson to the 
 TVPC $NN$ interaction are excluded, which is obviously true for the TVPC $\bar NN$ interaction 
 as well.

 \subsection{TVPC amplitude of ${\bar p}d$ forward scattering}
\label{widetildeg}
  One can write the ${\bar p}d$ forward elastic scattering amplitude in general form 
  taking into account the TVPC $\bar N N$ interactions, as it was done for
  $pd$ elastic scattering \cite{TUZyaf,UZTemPRC}, and then apply
  the generalized optical theorem to derive Eq.~(\ref{totalspin}) for
  the total ${\bar p}d$ scattering cross section.
  As in Ref. \cite{UZTemPRC}, the integrated cross section ${\widetilde \sigma}$
 is related to the TVPC term
 $\widetilde g$ of the ${\bar p}d$ forward elastic scattering amplitude by 
${\widetilde\sigma}=-4\sqrt{\pi}\,{\rm Im}\frac{2}{3}{\widetilde g}$.
Furthermore, the TVPC forward amplitude of $\bar p d$ elastic 
scattering $\widetilde g$ can be found within the Glauber theory
 \cite{UZTemPRC}.  
We  consider the $h_N$- and $g_N$-terms 
 and take into account both the $S$- and  $D$-wave components of the deuteron. Taking 
 into account that  the $g_N$-term is excluded
 in the process $\bar p n\to \bar p n$ due to the isospin operator in Eq.~(\ref{TVbNN}),
 we obtain the following result
 for the TVPC forward amplitude from the corresponding equation in Ref.~\cite{UZJHPRC16}:
 \begin{eqnarray}
\label{g5}
{\widetilde g}=\frac{i}{4{\pi}m_p}
\int_0^\infty dq q^2 \Bigl[S_0^{(0)}(q)-\sqrt{8} S_2^{(1)}(q) -4 S_0^{(2)}(q)+ \sqrt{2}\frac{4}{3} S_2^{(2)}(q)+
9 S_1^{(2)}(q)\Bigr]\\ \nonumber 
[-C^\prime_n(q)(h_p+g_p)-C^\prime_p(q)h_n] \ .
 \end{eqnarray}
Here $S_i^{(j)}$ 
are the  elastic form factors of the deuteron defined in Ref.~\cite{UZJHPRC16}. 
The first term in the (big) squared brackets in Eq.~(\ref{g5}), $S_0^{(0)}(q)$,  
 corresponds to the $S$-wave approximation, the second term, $S_2^{(1)}(q)$,
 accounts for the  $S$-$D$ interference,
and the last three terms contain the pure $D$-wave contributions.
 
As was shown in Ref.~\cite{UZJHPRC16}, the contribution of the $g'$-term to the null-test
signal vanishes in $pd$ scattering due to the specific spin-isospin structure of the $g'$-interaction.
Formally, for the same reason the charge-exchange  $g'$-term given by Eq. (\ref{chargex})
vanishes in the $\bar p d$ forward elastic scattering amplitude.

 In the first theoretical work \cite{Beyer} where the
 null-test signal was calculated within the impulse approximation, 
 the Coulomb interaction  was not considered.
 In Ref. \cite{Lazauskas} Faddeev calculations were performed, but only for $nd$ scattering
and at rather low energies of $\sim 100 $ keV. 
The Coulomb interaction was taken into account for the first time in Ref.~\cite{UZTemPRC} 
in a calculation of the null-test signal of $pd$ scattering within Glauber theory
 and found to be negligible. A similar result was found  
 in Ref.~\cite{LazGud2016} using Faddeev calculations.
  
\begin{figure}
\centering
\includegraphics[width=10cm,clip]{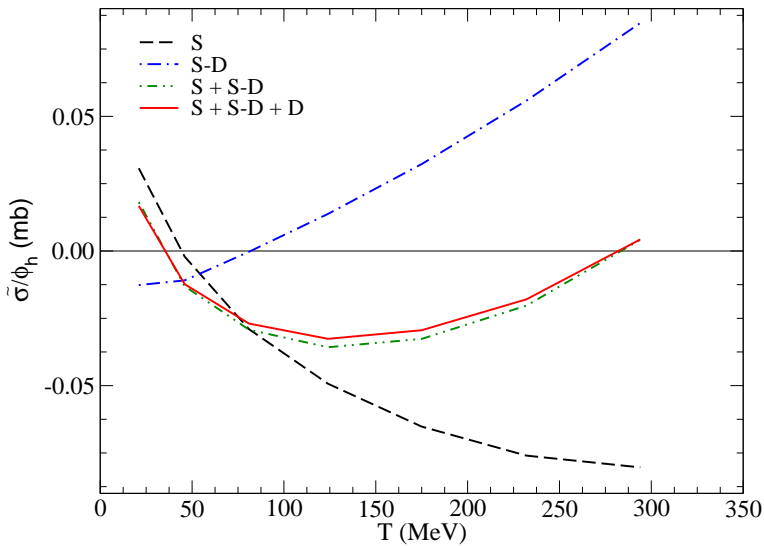}
\caption{The TVPC signal $\widetilde\sigma$ for the $h$-term, in units of
the ratio ($\phi_h$, see Ref.~\cite{UZTemPRC}) of the TVPC and the strong $h_1NN$ coupling constants, 
versus the antiproton beam energy $T$.
Results of our calculations accounting for different terms of the deuteron wave
function in Eq.~(\ref{g5}) are shown, based on the deuteron wave function of the CD Bonn potential 
and the hadronic $\bar p N$ amplitudes from Ref.~\cite{Nijmegen}:
$S$-wave (black), $S$-$D$ interference (blue), $S$ + $S$-$D$ waves (green), full result (red). 
}
\label{fig1}       
\end{figure}
 
 \subsection{Numerical results}
\label{numresults}
 Results of numerical calculations of the energy dependence of the null test-signal
 for the $h$-term are presented in Fig.~\ref{fig1}, in units of the unknown TVPC coupling
 strength. One can see from this figure that the deuteron
 $S$-wave contribution (dashed line) leads to a smooth energy dependence and has a node at
 an antiproton beam energy of about $50$ MeV. 
The inclusion of the $D$-wave changes this 
 behaviour considerably (solid line) due to a destructive $S$-$D$ interference (cf. dash-dotted line). 
 As a result, a second zero of the null-test signal $\widetilde {\sigma}$ appears at higher 
 energies, i.e. at $T\approx 300$ MeV. The maximal value
 of $\widetilde {\sigma}$ is expected at $100-150$ MeV. Note that the actual position of the 
 nodes changes only slightly when deuteron wave functions from other $NN$ 
 models are used for the calculation.

 Let us consider  possible spurious effects that could mimic a TVPC signal.
 One source for a spurious signal is associated  with a nonzero deuteron vector polarization 
 $p_d^y\not =0$ (in the direction of the incident-proton-beam polarization ${\bf p}^p$).
 In this case, the term  $\sigma_1P_y^{\bar p}p_y^d$ in Eq.~(\ref{totalspin}) contributes to the 
 asymmetry corresponding  to the difference  of the event counting rates for the cases
 of $p_y^{\bar p}P_{xz}>0$ and $p_y^pP_{xz}<0$ (with the fixed sign of $P_{xz}$), which is planned to be
 measured  at COSY \cite{TRIC}.
 According to our calculations, the integrated cross section $\sigma_1$ could be
 equal to zero at 
 antiproton beam energies of $\sim 100$~MeV (see results
 for the J\"ulich $\bar NN$ interaction model in Refs.~\cite{YUJH-PRC87,YUJH-PRC88}).
 Therefore, at this energy the spurios
 signal caused by a nonzero value of the deuteron vector polariziation $p_y^d$ could be minimized.
 
\section{Concluding remarks}
\label{conclusion} 

We have performed a study of time-reversal-invariance violating but parity conserving effects  
in antiproton-deuteron scattering. Specifically, we have evaluated the null-test TVPC signal
for scattering of antiprotons with transversal polarization $p_y^p$ on deuterons with tensor 
polarization $P_{xz}$ on the basis of the spin-dependent Glauber theory. 
The observed effects turned out to be similar to those in $pd$ scattering:
(i) There is a strong impact of the deuteron $D$-wave on the null-test signal
that arises from a destructive interference between the $S$- and $D$-wave contributions;
(ii) There is an oscillating behaviour of the null-test signal as a function of the beam energy.
Accordingly, it is possible that the signal for TVPC effects is zero at some 
specific energies, even when the TVPC interaction itself is nonzero.

\vskip 0.1cm 
{\bf Acknowledgement.} This work was supported in part by the Heisenberg-Landau program.

%
%
%

\end{document}